\documentclass[letterpaper]{article}
\pdfoutput=1 

\usepackage{jheppub} 

\usepackage[T1]{fontenc} 

\usepackage{graphicx} 
\usepackage{dcolumn} 
\usepackage{bm} 
\usepackage{amssymb} 
\usepackage{amsmath} 
\usepackage{amsfonts}
\usepackage{color}
\usepackage{comment}
\usepackage{rotating}
\usepackage{multirow}
\usepackage{colortbl}
\usepackage{afterpage}
\usepackage{slashed}
\usepackage{array}
\usepackage{subcaption}
\usepackage{bbm}

\usepackage[greek,english]{babel}
\newcommand{\cpi}{\text{\greektext p}}

\newcommand{\iu}{{\mathrm i}}
\newcommand{\E}{{\mathrm e}}

\def\PA#1{{\bf \color{red} [Prateek: #1]}}

\usepackage{tikz}
\usetikzlibrary{trees}
\usetikzlibrary{decorations.pathmorphing}
\usetikzlibrary{decorations.markings}

\usepackage{ctable}

\tikzset{
    photon/.style={decorate, decoration={snake}, draw=black},
    fermion/.style={draw=black, postaction={decorate},
        decoration={markings,mark=at position .55 with {\arrow[draw=black]{>}}}},
    scalar/.style={draw=black, dashed,postaction={decorate},
        decoration={markings,mark=at position .55 with {\arrow[draw=black]{>}}}},
    gluon/.style={decorate, draw=black,
        decoration={coil,amplitude=4pt, segment length=5pt}}
}

\title{Systematizing the Effective Theory of Self-Interacting Dark Matter}
\author{Prateek Agrawal, Aditya Parikh, and Matthew Reece}
\affiliation{Department of Physics, Harvard University, Cambridge, MA 02138, USA}
\date{today}

\abstract{
If dark matter has strong self-interactions, future astrophysical and cosmological observations, together with a clearer understanding of baryonic feedback effects, might be used to extract the velocity dependence of the dark matter scattering rate. To interpret such data, we should understand what predictions for this quantity are made by various models of the underlying particle nature of dark matter. In this paper, we systematically compute this function for fermionic dark matter with light bosonic mediators of vector, scalar, axial vector, and pseudoscalar type. We do this by matching to the nonrelativistic effective theory of self-interacting dark matter and then computing the spin-averaged viscosity cross section nonperturbatively by solving the Schr\"odinger equation, thus accounting for any possible Sommerfeld enhancement of the low-velocity cross section. In the pseudoscalar case, this requires a coupled-channel analysis of different angular momentum modes. We find, contrary to some earlier analyses, that nonrelativistic effects only provide a significant enhancement for the cases of light scalar and vector mediators. Scattering from light pseudoscalar and  axial vector mediators is well described by tree-level quantum field theory.
}

\begin{document}

\maketitle

\section{Introduction}
\label{Introduction}

The majority of mass in our universe is in the form of dark matter, but the underlying nature of dark matter and its interactions remains elusive. Many attempts to understand the nature of dark matter rely on searching for its interactions with ordinary matter, through direct or indirect detection experiments or attempts to directly produce dark matter particles at colliders or fixed-target experiments. We may also be able to learn about the particle nature of dark matter if it has significant self-interactions, which can alter the astrophysical and cosmological signals of dark matter in ways that may be detectable \cite{Spergel:1999mh, Dave:2000ar}. In recent years, many aspects of the astrophysics and cosmology of self-interacting dark matter have been extensively studied, so we will of necessity refer to only a small fraction of the literature. An excellent recent review article with extensive references is \cite{Tulin:2017ara}.

Dark matter self-interactions have been posited as a possible explanation for a number of discrepancies on small scales between observation and the results of classic $N$-body dark matter simulations of standard $\Lambda$CDM cosmology, such as the core/cusp problem \cite{Flores:1994gz, Moore:1994yx}, the missing satellites problem \cite{Moore:1999nt, Klypin:1999uc}, the too-big-to-fail problem \cite{BoylanKolchin:2011de}, and the diversity problem \cite{Oman:2015xda, deNaray:2009xj}. A weakness of such arguments is that baryonic physics, such as supernovae or AGN activity, can alter the distribution of dark matter in galaxies and clusters in ways that are unaccounted for in dark matter-only simulations. Incorporating baryonic physics in simulations accurately is challenging, and may solve some or even all of the potential problems with $\Lambda$CDM on small scales \cite{Governato:2009bg, Brooks:2012vi, Sawala:2015cdf,Brooks:2017rfe, Read:2018fxs, DelPopolo:2018wrz}.

In this paper, we do not take sides in this debate. We expect that in the future our understanding of dark matter and baryonic effects could be refined to allow for an averaged dark matter scattering rate to be extracted from data in systems on a variety of scales, from dwarf galaxies to clusters. Because the typical velocity of a dark matter particle is much lower in a dwarf galaxy than in a galaxy cluster, such observations could map out the velocity dependence of dark matter scattering, which encodes information about the underlying particle physics \cite{Feng:2009hw, Buckley:2009in, Loeb:2010gj}. An attempt to do this with current data has been made in \cite{Kaplinghat:2015aga} and argued to fit a Yukawa potential with a light mediator. Again, one can debate whether these conclusions are robust to uncertainties in baryonic physics, but in any case an important message to take away from \cite{Kaplinghat:2015aga} is that observations may eventually tell us the quantity $\langle \sigma v \rangle/m_{\rm DM}$ as a function of the rms velocity $v$.

With such measurements in hand, we would naturally want to answer the inverse problem: what underlying model of dark matter self-interactions produces the observed velocity dependence of the cross section? This question is nontrivial because dark matter is nonrelativistic, and in the nonrelativistic limit low-velocity scattering can be nonperturbatively enhanced by the Sommerfeld effect \cite{sommerfeld1931beugung}. To compute the cross section in the low-velocity regime, we must match the relativistic QFT of interest onto a nonrelativistic effective theory \cite{Caswell:1985ui,Lepage:1997cs} in which we can compute the cross section nonperturbatively by (numerically) solving the Schr\"odinger equation. In the dark matter context, the need for such calculations was first appreciated in the annihilation of heavy WIMPs \cite{Hisano:2002fk,Hisano:2003ec,Hisano:2004ds}. More recently, it has been applied to scattering of dark matter through light mediators that generate an effective Yukawa potential \cite{ArkaniHamed:2008qn,Buckley:2009in,Loeb:2010gj,Tulin:2013teo}. However, relativistic QFTs can match to a much wider range of nonrelativistic effective interactions than a simple Yukawa potential. The classification of such interactions \cite{Dobrescu:2006au} has been used to study a variety of possible recoil spectra in direct detection experiments \cite{Fan:2010gt, Fitzpatrick:2012ix, Bishara:2016hek} (see also \cite{Agrawal:2010fh,Freytsis:2010ne}). Most relevantly for our current work, it has also been used to classify dark matter self-interactions in \cite{Bellazzini:2013foa}, the ``effective theory of self-interacting dark matter.'' This paper is closely related to that one, although some of our conclusions differ.

\subsection{Goals of this work and relation to the previous literature}

Our goal in this paper is to calculate the velocity dependence of dark matter self-interactions for the case of spin-$1/2$ dark matter interacting via a light boson, which may be a scalar, pseudoscalar, vector, or axial vector. In each case, we match to a nonrelativistic effective theory, then solve the Schr\"odinger equation numerically to obtain the velocity dependence of the cross section.

Our work differs from, and extends, earlier work on nonrelativistic dark matter scattering in several respects. The first is our matching procedure: we match the Born approximation to short-distance scattering in the quantum mechanical effective theory to the tree-level perturbative QFT approximation to short-distance scattering. This provides a boundary condition at small radius, from which we can integrate outwards to solve the Schr\"odinger equation and capture long-distance effects of light mediators. The specifics of our matching procedure are inspired by earlier work \cite{Caswell:1985ui,Lepage:1997cs}, but the details are novel: our procedure is streamlined and easy to apply. A particular difference from some previous work on self-interacting dark matter is that we do not just consider the effective potential generated by $t$-channel exchange, but include the contact interactions arising in other channels. In the case of pseudoscalar mediators, this is crucial to obtain the correct cross sections at small velocities.

Our approach clarifies a number of issues related to pseudoscalar mediators. We sum over all angular momentum partial waves. While a simple Yukawa potential conserves orbital angular momentum, the exchange of a pseudoscalar or axial vector leads to interactions that couple different $l$ modes (while conserving the total angular momentum $j$). Some earlier work has considered a toy model aimed at approximating pseudoscalar exchange without treating the coupled channels carefully. This toy model includes a singular $1/r^3$ potential. We will show that this misses an important aspect of the physics, namely that the Born approximation for the pseudoscalar potential is non-singular at short distances. Furthermore,  our matching procedure is robust to variations in the matching scale. Unlike some results in the literature, our cross section is entirely determined by the underlying QFT and does not depend on ad hoc constants introduced in the nonrelativistic effective theory.

In the end, we find that the correct matching of a weakly-coupled effective field theory with a light pseudoscalar or axial vector mediator leads to a nonrelativistic effective theory in which there is no enhancement of the cross section at low velocities. For these theories, unlike the case of light scalar or vector mediators, tree-level QFT is reliable. This simple result is in contrast with some earlier claims in the literature, for instance, in studies of annihilating dark matter with a pseudoscalar mediator \cite{Bedaque:2009ri,  Liu:2013vha}. A similar claim about the lack of Sommerfeld enhancement for the pseudoscalar potential was made in \cite{Kahlhoefer:2017umn}. In this work, we provide more detailed arguments in support of this claim, incorporating both analytical and numerical evidence. 

We proceed as follows: in \S\ref{Explicit_Examples}, we introduce the basic models of mediators in QFT. We review why the axial vector mediator is special, in that its couplings to dark matter must vanish as the mediator mass goes to zero. In \S\ref{gen_procedure}, we describe our matching procedure, the way in which we set boundary conditions, and the process of extracting the $S$-matrix from numerical solutions of the Schr\"odinger equation. In \S\ref{Sommerfeld_enhancement}, we explain the absence of a Sommerfeld enhancement for a pseudoscalar  mediator, arguing that a 1-loop calculation in the perturbative QFT also suggests that the effect should be absent. We explain the differences between our results and certain claims in the literature. Then, we present the numerical results of solving the Schr\"odinger equation and computing the cross section  for the various models. We offer concluding remarks in \S\ref{sec:conclusions}.

\section{SIDM and Our Examples}
\label{Explicit_Examples}

We consider a weakly coupled dark sector with spin-$1/2$ fermionic dark matter interacting via scalar or vector mediators. If these mediators are light and generate an attractive potential, then it is possible that Sommerfeld enhancement can significantly boost the scattering cross section. So, it is crucial to analyze the potentials generated by various renormalizable interactions. 

The details of the calculation depend on whether the dark matter carries an approximate conserved charge. If dark matter is a Majorana fermion, then the $\chi \chi \to \chi \chi$ scattering process receives contributions from $s$-, $t$-, and $u$-channel diagrams. In the Majorana case, there is no vector interaction, i.e., the coupling $A_\mu \chi^\dagger \overline{\sigma}^\mu \chi$ takes the form of an axial vector interaction when packaged into a 4-component field. If dark matter carries an approximately conserved charge, we should consider a Dirac fermion, and the dark matter abundance may be primarily of one charge ({\em asymmetric} dark matter  \cite{Nussinov:1985xr,Kaplan:1991ah,Zurek:2013wia}) or it may contain particles of both charges ({\em symmetric} dark matter). Scattering in the asymmetric case, $\chi \chi \to \chi \chi$, receives contributions from only $t$- and $u$-channel diagrams. In the symmetric case, there is additionally the process $\chi \overline{\chi} \to \chi \overline{\chi}$, which receives $s$- and $t$-channel contributions but no $u$-channel contribution. 

Our goal in this work is to clarify conceptual issues in matching theories of self-interacting dark matter to a nonrelativistic effective theory, and to understand in which cases a Sommerfeld enhancement is present. These aspects of the physics are not sensitive to the Majorana or Dirac nature of the fermion or to the symmetric or asymmetric nature of the dark matter population. Thus, for concreteness, in the remainder of the paper we will only discuss the case of Dirac dark matter and $\chi \overline{\chi} \to \chi \overline{\chi}$ scattering. All of our results can be straightforwardly generalized to the other cases.

We denote the dark matter field as $\chi$. The consistency of a Dirac effective theory requires a good approximate global symmetry,
\begin{align}
    \chi \rightarrow \text{e}^{\iu \alpha}\chi 
\end{align}
In the absence of such a symmetry, we could write Majorana mass terms which split the Dirac fermion into two Majorana mass eigenstates. 

We now list various cases that we will individually consider in this paper. The Lagrangian for dark matter coupling with a real scalar $\phi$ is,
\begin{align}
    \mathcal{L}_{\text{scalar}} = \iu\overline{\chi}\gamma^{\mu}\partial_{\mu}\chi - m_{\chi}\overline{\chi}\chi + \frac{1}{2}\partial_{\mu}\phi\partial^{\mu}\phi - \frac{1}{2}m_{\phi}\phi^{2} - \lambda\phi\overline{\chi}\chi.
\end{align}
If instead we wish to couple to a pseudoscalar $\phi$, we obtain the following,
\begin{align}
    \mathcal{L}_{\text{pseudoscalar}} = \iu\overline{\chi}\gamma^{\mu}\partial_{\mu}\chi - m_{\chi}\overline{\chi}\chi + \frac{1}{2}\partial_{\mu}\phi\partial^{\mu}\phi - \frac{1}{2}m_{\phi}\phi^{2} - \iu\lambda\phi\overline{\chi}\gamma^{5}\chi.
\end{align}
Note that we have assumed a $\sf{CP}$ symmetry to a good approximation, so that $\phi$ has well-defined $\sf{CP}$ quantum numbers.

We can also couple the dark matter to a U(1) gauge field $\phi_{\mu}$. If the charge assignments are vectorlike (i.e.~we simply gauge the Dirac U(1) symmetry), then we end up with the vector interaction of the gauge field with the dark matter.
In order to generate a mass for the U(1) gauge field, we can couple it to a charged scalar that gets a vev. Imposing a separate global U(1) Dirac symmetry under which the scalar is a singlet forbids a coupling of this scalar to the dark matter.
\begin{align}
    \mathcal{L}_{\text{vector}} 
    &= 
    \iu\overline{\chi}\gamma^{\mu}\partial_{\mu}\chi - m_{\chi}\overline{\chi}\chi - \frac{1}{4}F_{\mu\nu}F^{\mu\nu} 
    + \frac{1}{2}m_{\phi}^{2}\phi_{\mu}\phi^{\mu} 
    - \lambda \phi^{\mu}
    \overline{\chi} \gamma_{\mu}\chi
    \,.
\end{align}

To obtain a purely axial vector interaction of Dirac fermion dark matter is a little intricate.  For clarity, we start with two left-handed Weyl fermions,
$\chi_1$ and $\chi_2$, with the same charge under the U(1) gauge group. In this case, the Dirac mass term is not invariant under a U(1) gauge transformation, and must arise from the U(1) breaking. This also implies that the coupling of an axial vector mediator to a massive fermion must vanish in the limit that the vector boson mass goes to zero. This distinguishes the axial vector case from the other cases, in which we are free to take the couplings to be order-one numbers. Explicitly, we write an abelian Higgs model with a Higgs boson of charge 2:\footnote{One could consider other possibilities, e.g., a Higgs boson of charge 1 that couples quadratically to give the fermion a mass through a higher-dimension operator. This would only make the problem of achieving a large coupling for a light mediator more severe.}
\begin{align}
    \mathcal{L}_{\text{axial vector}} 
    &=  
    - \frac{1}{4}F_{\mu\nu}F^{\mu\nu} 
    + |(\partial_{\mu} - 2\iu\lambda\phi_{\mu})H|^{2}
    + \frac{\lambda_{q}}{2} (H^{\dagger}H - v^2)^2
    \nonumber\\ &\qquad
    + \iu\chi_{1}^{\dagger}\overline{\sigma}^{\mu}\partial_{\mu}\chi_{1} 
    + \iu\chi_{2}^\dagger \overline{\sigma}^{\mu}\partial_{\mu}\chi_{2} 
    - \lambda\phi^{\mu}\chi_{1}^{\dagger}\overline{\sigma}_{\mu}\chi_{1} 
    - \lambda\phi^{\mu}\chi_{2}^\dagger
    \overline{\sigma}_{\mu}\chi_{2}
    - [yH\chi_{1}\chi_{2} + \mathrm{h.c.}] 
    \,.
\end{align}
Again, an additional global U(1) Dirac symmetry needs to be imposed to ensure that terms like $H \psi_1 \psi_1$ are absent.
After SSB, we can expand the Higgs around its expectation value: $H = v + \frac{1}{\sqrt{2}}(h + \iu \phi_{2})$. Keeping the relevant terms, we find in unitary gauge,
\begin{align}
    \mathcal{L}_{\text{axial vector}} 
    &=  
    - \frac{1}{4}F_{\mu\nu}F^{\mu\nu} + 4\lambda^{2}v^{2}\phi_{\mu}\phi^{\mu} \nonumber \\&\qquad
    + \frac{1}{2}|\partial_{\mu}h|^{2}
    - \lambda_q v^2 h^{2} 
    \nonumber \\ & \qquad
    +\iu\overline{\chi}\gamma^{\mu}\partial_{\mu}\chi - yv\overline{\chi}\chi 
    + \lambda\phi^{\mu}\overline{\chi}\gamma_{\mu}\gamma^{5}\chi 
    - \frac{y}{\sqrt{2}}h\overline{\chi}\chi 
\end{align}
The masses and couplings of the massive vector, radial Higgs mode, and fermion are given in terms of the fundamental parameters as,
\begin{equation}
    m_{\phi}^{2} = 8\lambda^{2}v^{2} \quad \quad m_{h}^{2} = 2\lambda_q v^2 \quad \quad m_{\chi} = yv .
\end{equation}
In particular, this scenario predicts that the coupling of the axial vector mediator to the dark matter behaves as 
\begin{equation}
    \lambda = \frac{y}{2\sqrt{2}} \frac{m_\phi}{m_\chi},
\end{equation} 
so light axial-vector mediators  are necessarily weakly coupled.

From an effective field theory point of view, all these interactions are equally well motivated to analyze. Furthermore, if we keep an eye towards UV completions, some of these scenarios arise more naturally than others. Light scalars that are not pseudo-Nambu-Goldstone bosons are unnatural. Light vectors are also unnatural, if their mass comes from Higgsing, because the Higgs boson mass must also be protected. On the other hand, pseudoscalars are particularly well motivated because we know examples of underlying dynamics which can generate pseudoscalar couplings with an associated light boson, such as pions in QCD. So it is much easier to embed a light pseudoscalar into a UV complete theory. 

\section{General Procedure}
\label{gen_procedure}

In this section, we outline the general procedure of obtaining the scattering cross sections, starting from a weakly-coupled Lagrangian. For light mediators, there can be substantial effects from multiple exchanges of the mediator in nonrelativistic scattering processes, so that the tree-level approximation does not reflect the true answer. A convenient way to resum these contributions in this case is to map the problem on to the equivalent quantum mechanical scattering problem. We can then solve the QM problem nonperturbatively and extract the scattering matrix elements. These amplitudes include the putative Sommerfeld enhancement effects.

Before we outline the procedure, it is worth highlighting the validity of the procedure carried out below. At small enough values of the coupling, the scattering amplitude at some fixed velocity is well-approximated by the tree-level amplitude in the QFT. If the relativistic corrections are small, the same amplitude is also well described by the Born approximation in a quantum mechanical system. The scattering potential is calculated by matching a QFT amplitude with the corresponding QM amplitude in the Born approximation. However, a subtlety that we will encounter involves potentials where the higher-order Born terms are ``divergent''. This is the case for singular potentials that grow faster than $1/r^2$ as $r\to 0$. In such a case, to calculate the higher-order Born terms we would need to regulate and renormalize the quantum mechanical problem, and consistently match with the QFT at loop level. 
Indeed, this is true for the potentials we consider, but these potentials do have a well-defined first order Born limit as $r\to 0$ which will be sufficient for us to avoid the issue of divergences.

Our problem neatly factorizes into two pieces. The short-distance $r\to 0$ piece is the part where the potential divergence shows up,  but from the QFT point of view this corresponds to high energy scattering, where Sommerfeld enhancement should not play a role and the amplitude should be well approximated by the tree-level diagram, or equivalently the first  Born approximation in the QM picture. The potential Sommerfeld enhancement from multiple exchanges of the mediator appears in the larger $r$ region, where the QM potential is well-behaved.

\begin{figure}[tp]
    \centering
    \includegraphics[width=0.75\textwidth]{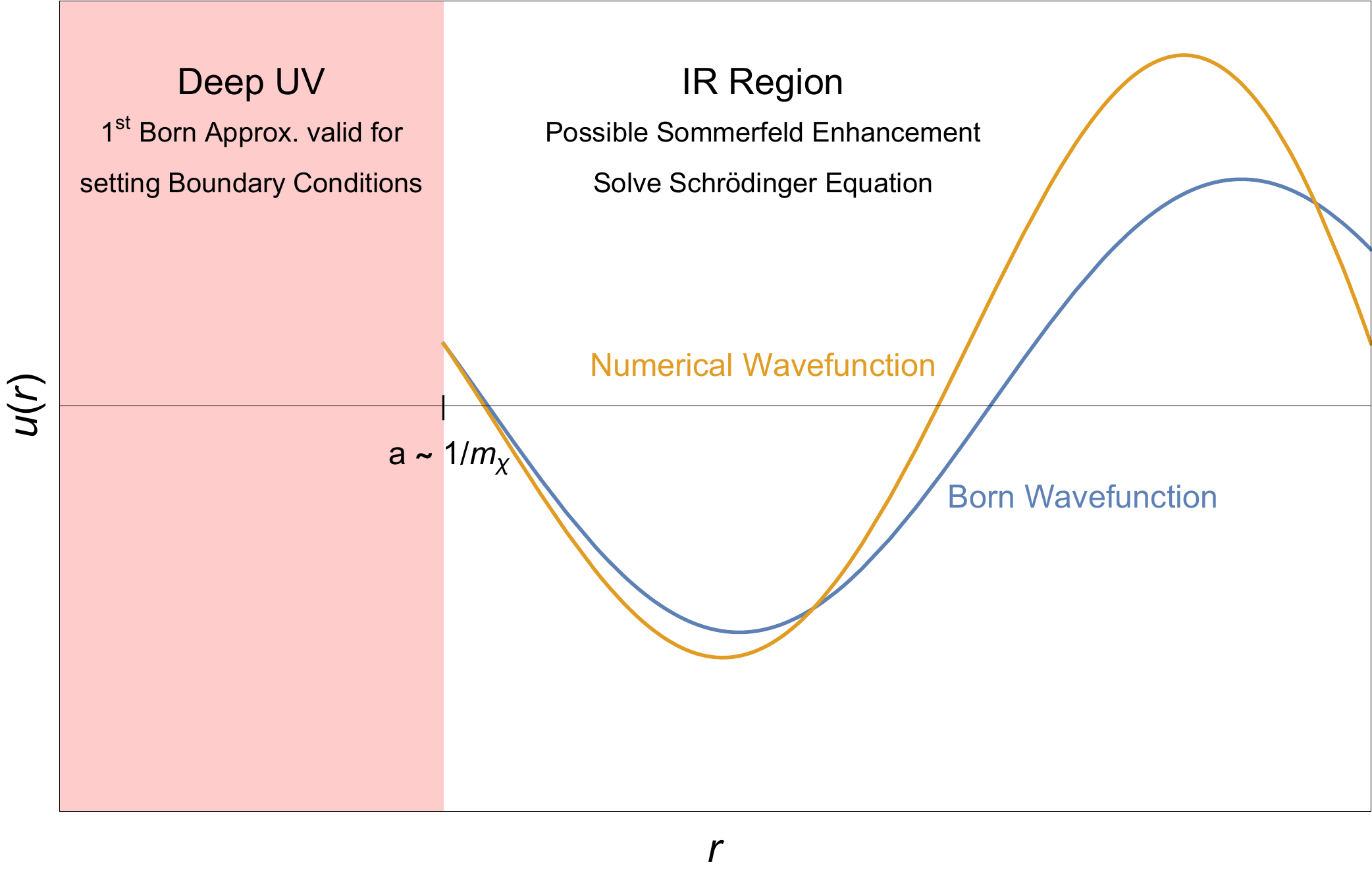}
    \caption{A schematic of our matching procedure. In the deep UV, we expect the Born approximation to hold, which we use to set our boundary conditions at $r = a$. Beyond this is the IR of our theory where Sommerfeld enhancement might be important and we need to solve the Schr\"odinger equation with the appropriate potential.}
    \label{fig:IC_Schematic}
\end{figure}

Thus, we use the following procedure, described in further detail below. We match the QFT tree-level amplitude with the QM amplitude in the first Born approximation to calculate the matching condition close to the origin at $r=a$. The nonrelativistic effective theory should not be expected to accurately describe momenta of order $m_\chi$, at which dark matter particles are semirelativistic. As a result, the natural radius at which to match the relativistic theory to the nonrelativistic theory is the Compton radius of the dark matter, $a \simeq 1/m_\chi$. With the boundary condition established by matching, we solve the Schr\"{o}dinger equation numerically, and thereby derive the scattering amplitude. We illustrate this procedure schematically in Figure $\ref{fig:IC_Schematic}$.
We then proceed to extract the S-matrix from the QM solution, highlighting the coupled channel case. We will then be in a position to critically evaluate which interactions give rise to a Sommerfeld enhancement. 

\label{sec:matching}

\subsection{Computing the Tree Level Potential}
To compute the tree level potential, we first compute the tree-level perturbative QFT amplitude for the process we are interested in. To illustrate our procedure, we consider  $\chi\overline{\chi}\rightarrow\chi\overline{\chi}$. At tree level, this has contributions from an $s$- and a $t$-channel Feynman diagram. Next, we take the nonrelativistic limit of this amplitude and keep the leading terms. The scattering amplitude in the Born approximation in nonrelativistic quantum mechanics is given by
\begin{equation}
    \langle \vec{p}_{\rm f} | \iu \mathcal{T} | \vec{p}_{\rm i} \rangle = -\iu \widetilde{V}(\vec{q})(2\cpi)\delta(E_{p_{\rm f}}-E_{p_{\rm i}}), \quad \quad \vec{q} = \vec{p}_{\rm f} - \vec{p}_{\rm i}.
\end{equation}
Comparing this expression to the nonrelativistic limit of our QFT amplitude gives us
\begin{equation}
    \widetilde{V}(\vec{q}) = -\frac{1}{4E_{p_{f}}E_{p_{i}}}M,
\end{equation}
where the extra factors of $2E_{p}$ come from the difference in the conventional normalizations of relativistic and nonrelativistic single particle states. Finally, to compute the real space potential $V(\vec{r})$, we have to Fourier transform $\widetilde{V}(\vec{q})$ with respect to $\vec{q}$.

We obtain the following potentials
\begin{align} 
  V_{\text{scalar}}(r) 
  &=  
  -\frac{\lambda^{2}}{4\cpi r}\E^{-m_{\phi}r}
  \\
V_{\text{pseudoscalar}}(r) 
&=  
\frac{\lambda^2}{4\cpi}
\left(\frac{4\cpi\delta^{3}(\vec{r})}{4m_{\chi}^{2}
- m_{\phi}^{2}}\Big(\frac{1}{2} - 2S_{1}\cdot S_{2}\Big) - \frac{4\cpi
\delta^{3}(\vec{r})}{3 m_{\chi}^{2}}\E^{-m_{\phi}r}S_{1}\cdot S_{2} 
\right.
\nonumber \\& \qquad 
\left.+ \frac{\E^{-m_{\phi}r}}{m_{\chi}^{2}}\Big[\frac{m_{\phi}^{2}}{3r}S_{1}\cdot S_{2} + \frac{3(S_{1}\cdot\hat{r})(S_{2}\cdot\hat{r}) - S_{1}\cdot S_{2}}{r^{3}}\Big(1 + m_{\phi}r + \frac{m^{2}_{\phi}r^{2}}{3}\Big)\Big]
\right)
\label{pseudoscalar_potential_equation}
\\
V_{\text{vector}}(r) &= \frac{\lambda^{2}\delta^{3}(\vec{r})}{4m_{\chi}^{2} - m_{\phi}^{2}}\Big(\frac{3}{2} + 2S_{1}\cdot S_{2}\Big) -\frac{\lambda^{2}}{4\cpi r}\E^{-m_{\phi}r}
\\
\label{axial_vector_potential_equation}
V_{\text{axial vector}}(r) 
&= \frac{\lambda^{2}\delta^{3}(\vec{r})}{4m_{\chi}^{2} - m_{\phi}^{2}}\Big(2S_{1}\cdot S_{2} - \frac{1}{2}\Big) -\frac{\lambda^{2}}{\cpi r}\E^{-m_{\phi}r}S_{1}\cdot S_{2} + \frac{4m_{\chi}^{2}}{m_{\phi}^{2}}V_{\text{pseudoscalar}}
\end{align}
The terms in these potentials arising from $t$-channel contributions have been previously computed (e.g.,  \cite{Moody:1984ba, Fan:2010gt, Bellazzini:2013foa, Daido:2017hsl}). To leading order in $q^{2}$, our results also agree with those in \cite{Fadeev:2018rfl}. The $s$-channel contributions provide contact terms that we have written in the form of a nonrelativistic potential via Fierz rearrangement (see, e.g., \cite{Agrawal:2010fh}). As detailed in Appendix~\ref{Spinor_Minus_Signs}, the spin matrices for antifermions come with an additional minus sign.

In the pseudoscalar potential, we observe two delta-function terms. The first term comes from the $s$-channel, but the second term comes from the $t$-channel, which can be seen in the following manner:
\begin{align}
\partial_{i}\partial_{j}\Big(\frac{\E^{-m_{\phi}r}}{r}\Big) & = 
 \E^{-m_{\phi}r}\Big[\frac{1}{3}\nabla^{2}\delta_{ij} + \partial_{i}\partial_{j} \Big]\frac{1}{r} + 2\Big[\hat{r}_{i}\frac{-1}{r^{2}}\hat{r}_{j}(-m_{\phi}\E^{-m_{\phi}r}) \Big] + \frac{1}{r}\partial_{i}\Big[-m_{\phi}\hat{r}_{j}\E^{-m_{\phi}r}\Big] \nonumber \\
\\ 
&= \E^{-m_{\phi}r}\frac{m_{\phi}^{2}}{3r}\Bigg[\Big(3\hat{r}_{i}\hat{r}_{j} - \delta_{ij}\Big)\Big(1 + \frac{3}{m_{\phi}r} + \frac{3}{m_{\phi}^{2}r^{2}}\Big) + \delta_{ij}\Bigg] - \frac{4\cpi}{3}\delta^{3}(r)\delta_{ij}\E^{-m_{\phi}r}.
\end{align}

\subsection{The Schr\"{o}dinger Equation}
\label{sec:setup}
We solve the Schr\"odinger equation in the partial wave expansion using the potential derived above. 
The rotational invariance of the Hamiltonian  implies that $j$ is a conserved quantum number. For the $2\to2$ scattering process involving spin-$\frac12$ particles that we consider, the total spin is $s=0,1$. Consequently, it will be convenient to work in a basis of states $|r,j,\sigma,\ell,s\rangle$, where $\ell$ and $s$ are the total orbital and spin angular momenta respectively, and $\sigma$ is the $z$-component of the total angular momentum. Some of the terms in the potentials we consider mix terms with different $\ell$ values, so for a given $j$, we get a $4\times 4$ block diagonal Hamiltonian. 
The wavefunction $\psi(\vec{r})$ can be separated,
\begin{align}
    \psi(\vec{r})
    &=
    \sum_{j,\sigma,\ell,s}
    \frac{u_{j\sigma\ell s}(r)}{r}
    \langle \hat{n} | j \sigma \ell s\rangle
    \,.
\end{align}
The potential preserves $j,\sigma$, so without loss of generality we can set $\sigma=0$. 
The Schr\"{o}dinger equation for a fixed value of $j$ becomes
\begin{align}
    \frac{1}{2\mu}\left(
    -\partial_r^2
    -k^2 
    + \frac{\ell(\ell+1)}{r^2}
    \right)
    u_{\ell s}(r)
    + \sum_{\ell's'} V_{\ell s,\ell's'}(r) u_{\ell' s'}(r)
    &=0
\end{align}
The matrix elements of the potential $V(r)$ are given in Appendix~\ref{Pseudoscalar_Clebsch_Gordan}. The subscripts $(\ell,s)$ take on 4 possible values $\{(j,0),(j-1,1),(j,1),(j+1,1) \}$. We have suppressed the $j$ and $\sigma$ quantum numbers for notational clarity. Conveniently, the four $(\ell, s)$ states separate into two states that mix with each other ($|j\pm 1, 1\rangle$) and two that evolve independently ($|j,0\rangle$ and $|j,1\rangle$).

Assuming that $rV(r)\rightarrow 0$ as $r\rightarrow\infty$,  the asymptotic wavefunction should be a solution to the free particle equation. The solutions of the radial free particle equation, denoted $s_{\ell}$ and $c_{\ell}$, are given in terms of the spherical Bessel functions.
\begin{align}
    s_{\ell}(kr) 
    &\equiv
    k r j_{\ell}(kr), 
    \qquad
    c_{\ell}(kr) 
    \equiv -k r y_{\ell}(kr)
    \, .
\end{align}
The time-independent solution to the Schr\"odinger equation 
can also be interpreted as a solution to a scattering problem. The asymptotic form of the solution is a combination of an incoming wave and a scattered wave. Different choices of boundary conditions correspond to different possible incoming waves -- there are as many independent boundary conditions as the number of equations. A convenient choice of basis is to label them by $(\ell s)$ values themselves.
The asymptotic solutions to the Schr\"odinger equation can be written in terms of the free particle solutions as,
\begin{align}
    u_{\ell s}^{(\ell' s')} (r) 
    &\sim 
    \delta_{\ell' s', \ell s} s_{\ell}(kr) 
    + K_{\ell' s', \ell s} \, c_{\ell}(kr)
    \,.
\end{align}
The $\mathbf{K}$-matrix is the generalization of the more familiar partial wave phase-shift $\tan \delta_l$. 
By solving the coupled differential equations numerically, we can extract the $\mathbf{K}$-matrix, and then calculate the scattering cross section. We describe these steps in detail next.

\subsection{Setting Boundary Conditions}
\label{Boundary_Conditions}

\begin{figure}[tp]
\centering
    \includegraphics[width=0.75\textwidth]{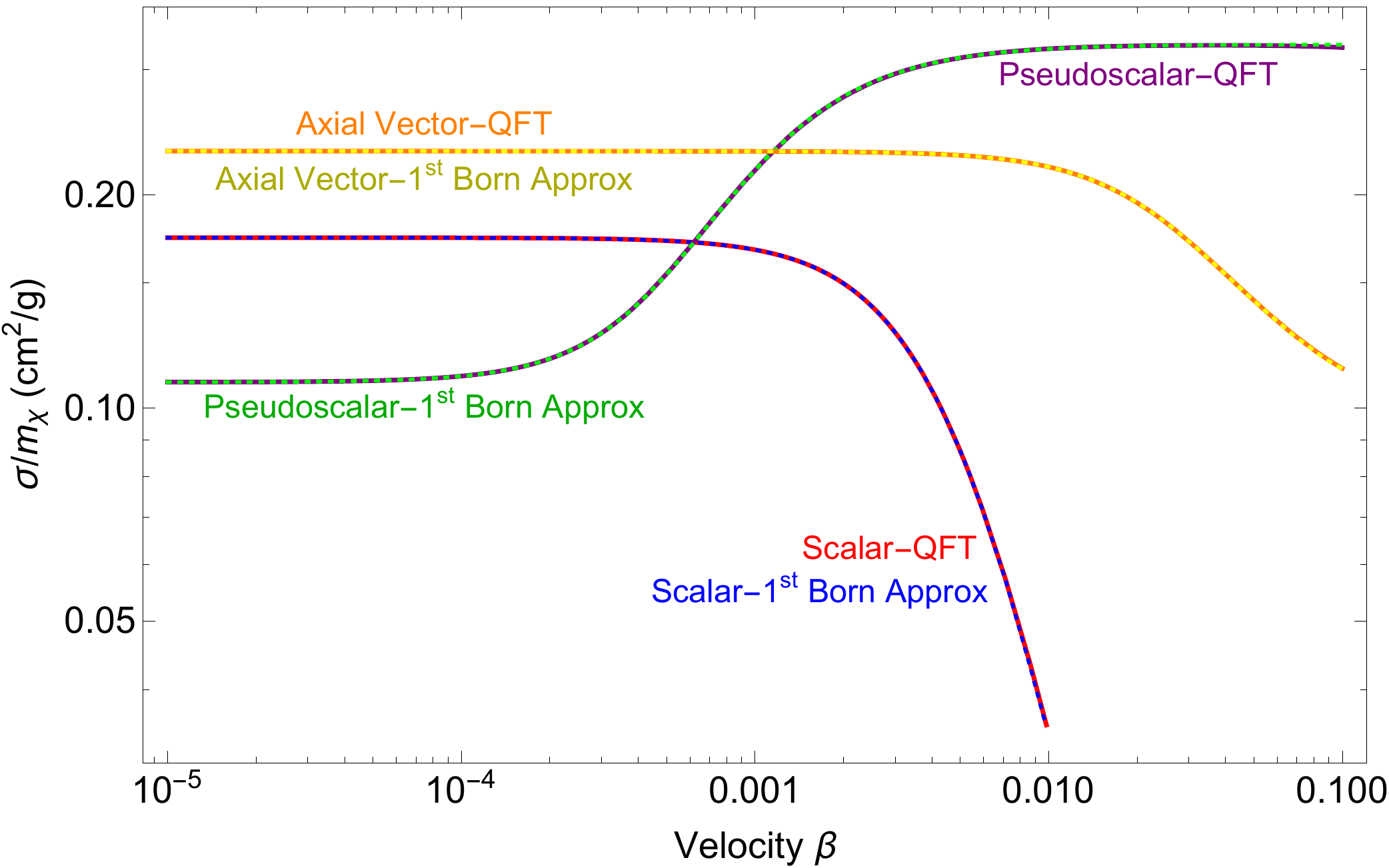}
\caption{Illustration of the validity of the matching procedure via numerical evaluation of the Born cross section compared to the tree-level QFT cross section as a function of velocity for the Yukawa (scalar) potential, pseudoscalar potential and axial vector potential. For the scalar case, we choose $\lambda = 10^{-1}$, $m_{\phi} = 10^{-2}$ GeV and $m_{\chi} = 1$ GeV. For the pseudoscalar case, we choose $\lambda = 10^{-1}$, $m_{\phi} = 10^{-3}$ MeV and $m_{\chi} = 1$ MeV. For the axial vector case, we work in the decoupling limit. We choose the vev $v = 20$ MeV, $\lambda_{\text{q}} = 4\cpi$, $\lambda = 10^{-3}$ and $y = 1$. We emphasize that this plot should {\em not} be taken as an accurate illustration of the full cross section, as the Born approximation receives large corrections in the Yukawa case.}
\label{fig:QFT_Born_comparison}
\end{figure}

The boundary conditions for the Schr\"{o}dinger equation above are set using the Born approximation in the region $r<a = m_{\chi}^{-1}$ as outlined in Section~\ref{sec:matching}. This region of small radii probe the UV of our effective quantum mechanical description and we expect this to match onto the corresponding QFT. Therefore, we expect the first Born approximation to reproduce the tree level perturbative QFT approximation. We show this matching in Figure~\ref{fig:QFT_Born_comparison}. The $\mathbf{K}$-matrix in the Born approximation is simple to calculate,
\begin{align}
\label{first_born_integrand}
    K_{\ell s, \ell's}^a
    &= 
    -\frac{2\mu}{k} \int_{0}^{a} {\rm d} r\,
    s_{\ell'}(kr) s_{\ell} (k r) 
    V_{\ell s,\ell' s'}(r)
\end{align}
where we have restricted the integral to $r<a$. This gives us our boundary conditions for the numerical solutions
\begin{align}
    u_{\ell s}^{(\ell' s')} (a) 
    &\sim 
    \delta_{\ell' s', \ell s} s_{\ell}(ka) 
    + K^a_{\ell' s', \ell s} \, c_{\ell}(ka)
    \,.
\end{align}
which we can use to evaluate the wavefunction and its derivative at $r = a = m_{\chi}^{-1}$. In our numerical calculations for the various potentials below, we have checked that varying the matching radius has little effect on our results, provided that we choose $a$ of order $m_{\chi}^{-1}$.

In this whole discussion so far, we have not specified the structure of the potential. In particular, potentials in quantum mechanics can diverge at the origin, and the Born limit may not be well-defined. A potential is considered non-singular if $r^{2}V(r)\rightarrow 0$ as $r\rightarrow 0$. Since $s_\ell(k r) \sim r^{l+1}$ for small $r$, for such potentials clearly the integral above is convergent. The pseudoscalar potential does not fit this criterion due to the $r^{-3}$ piece and naively looks singular when $\ell = \ell'= 0$. However, as we see from the results in Appendix \ref{Pseudoscalar_Clebsch_Gordan}, these dangerous terms vanish under the action of the operator $\mathcal{O}_{T} \equiv 3(\vec{S_{1}}\cdot\hat{r})(\vec{S_{2}}\cdot\hat{r}) - \vec{S}_{1}\cdot\vec{S}_{2}$ appearing in the numerator.

Since we consider all potential terms that are generated from tree-level exchanges, it is interesting to note that the QFT seems to produce highly non-generic potentials which may seem singular, but possess operator structures that remove these divergent pieces at the leading Born approximation.

\subsection{Extracting the S-Matrix from Numerical Solutions to the Schr\"odinger Equation}
The $\mathbf{K}$-matrix can be extracted from the set of asymptotic solutions $u^{(\ell' s')}_{\ell s}(r)$ at some suitably large value of $r = r_{\rm max}$. Define the matrix $\mathbf{W}$ in terms of the matrix of solutions $\mathbf{u}$,
\begin{align}
\textbf{W}
&=
\left(\mathbf{u}'(r_{\rm max})\right)^{-1}
\cdot \mathbf{u}(r_{\rm max})
\end{align}
with the $'$ denoting $\partial_r$.
The $\mathbf{K}$-matrix is
\begin{align}
    \mathbf{K}
    &=
    \left(\mathbf{s}(k r_{\rm max}) 
    - \mathbf{W}\cdot \mathbf{s}'(k r_{\rm max})\right)
    \cdot
    \left(\mathbf{c}'(k r_{\rm max})\cdot \mathbf{W}_{} -  \mathbf{c}(k r_{\rm max}))\right)^{-1}
\end{align}
where we have defined the free-particle solution matrices as
\begin{align}
    \mathbf{s}(kr) &= 
    \mathrm{diag}\left[s_j(kr),s_{j-1}(kr),s_{j}(kr),s_{j+1}(kr)\right],
    \nonumber\\
    \mathbf{c}(kr) &= 
    \mathrm{diag}\left[c_j(kr),c_{j-1}(kr),c_{j}(kr),c_{j+1}(kr)\right].
\end{align}
The $\mathbf{S}$-matrix follows,
\begin{align}
    \textbf{S} = (\mathbbm{1} + \text{i}\textbf{K})\cdot(\mathbbm {1} - \text{i}\textbf{K})^{-1}
    \,.
\end{align}
We outline the steps involved in obtaining the cross section from the S-matrix in this basis following the notation in Ref.~\cite{Weinberg:1995mt}.
The differential cross section is given in terms of the scattering amplitude in the familiar way,
\begin{align}
    \frac{d\sigma\left( \hat{k},\sigma_1,\sigma_2 \to \hat{k}',\sigma_1',\sigma_2'\right)}{d\Omega}
    &=
    \left|f\left( \hat{k},\sigma_1,\sigma_2 \to \hat{k}',\sigma_1',\sigma_2'\right)\right|^2
    \,.
\end{align}
We can write $f$ in terms of the  $(j\sigma \ell s)$ basis S-matrix using the Wigner-Eckart theorem,
\begin{align}
    f\left( \hat{k},\sigma_1,\sigma_2 \to \hat{k}',\sigma_1',\sigma_2'\right)
&= -\frac{2\cpi \iu}{k}
\sum_{j \sigma l' s' p' l s p}
C_{\frac12 \frac12} (s, p; \sigma_1 \sigma_2) 
C_{l s} (j, \sigma; m, p) 
\nonumber\\&\qquad\qquad\qquad\qquad\times
C_{\frac12 \frac12} (s',p' ; \sigma_1' \sigma_2') 
C_{l' s'} (j, \sigma; m', p') 
\nonumber\\&\qquad\qquad\qquad\qquad\times
Y_{l}^{m *}(\hat{k}) 
Y_{l'}^{m'}(\hat{k}') 
 \left(\mathbf{S}^{j}(E)-1\right)_{l's';ls}
 \,,
\end{align}
where the $C$'s are Clebsch-Gordan coefficients. We have specialized to the case of elastic $2\to 2$ scattering between spin-$\frac12$ particles. The spin-averaged cross section in this basis is,
\begin{align}
\sigma
&= \frac{\cpi}{4k^{2}}
\sum_{j,\ell's',\ell s}(2j+1)
\left|\left(
{\mathbf S}^{j}(E)-\mathbbm{1}
\right)_{\ell's',\ell s} 
\right|^{2}
\,.
\end{align}
In SIDM phenomenology, there are other related quantities of interest, such as the momentum transfer cross section or the viscosity cross section. Their utility arises from their well-behaved soft and forward limits, and they have a direct bearing on the evolution of dark matter phase space in halos. They are defined in terms of the differential cross section. The transfer cross section\footnote{See \cite{Kahlhoefer:2013dca,Agrawal:2016quu} for further discussion.} is defined as \cite{Feng:2009hw, Buckley:2009in,Tulin:2013teo}
\begin{equation}
\sigma_{T} = \int \frac{{\rm d}\sigma}{{\rm d}\Omega} (1 - \cos\theta) {\rm d}\Omega 
\end{equation}
while the viscosity cross section is defined as 
\begin{align}
\sigma_{V} 
&= 
\int \frac{{\rm d}\sigma}{{\rm d}\Omega} (1 - \cos^{2}\theta) {\rm  d}\Omega 
\,.
\end{align}

\section{Sommerfeld Enhancement}
\label{Sommerfeld_enhancement}

In this section we study the existence of Sommerfeld enhancement for
various types of interactions. We begin with a discussion of our
results for the Yukawa potential, which has previously been studied
extensively
\cite{ArkaniHamed:2008qn,Buckley:2009in,Loeb:2010gj,Tulin:2013teo}. We
then present novel results for the pseudoscalar and axial vector
cases, which do not show Sommerfeld enhancement. We discuss a
diagrammatic argument that reinforces this conclusion, and comment on
disagreement with previous work that has found Sommerfeld enhancement
in the pseudoscalar case.

\begin{figure}[tp]
    \centering
    \includegraphics[width=0.75\textwidth]{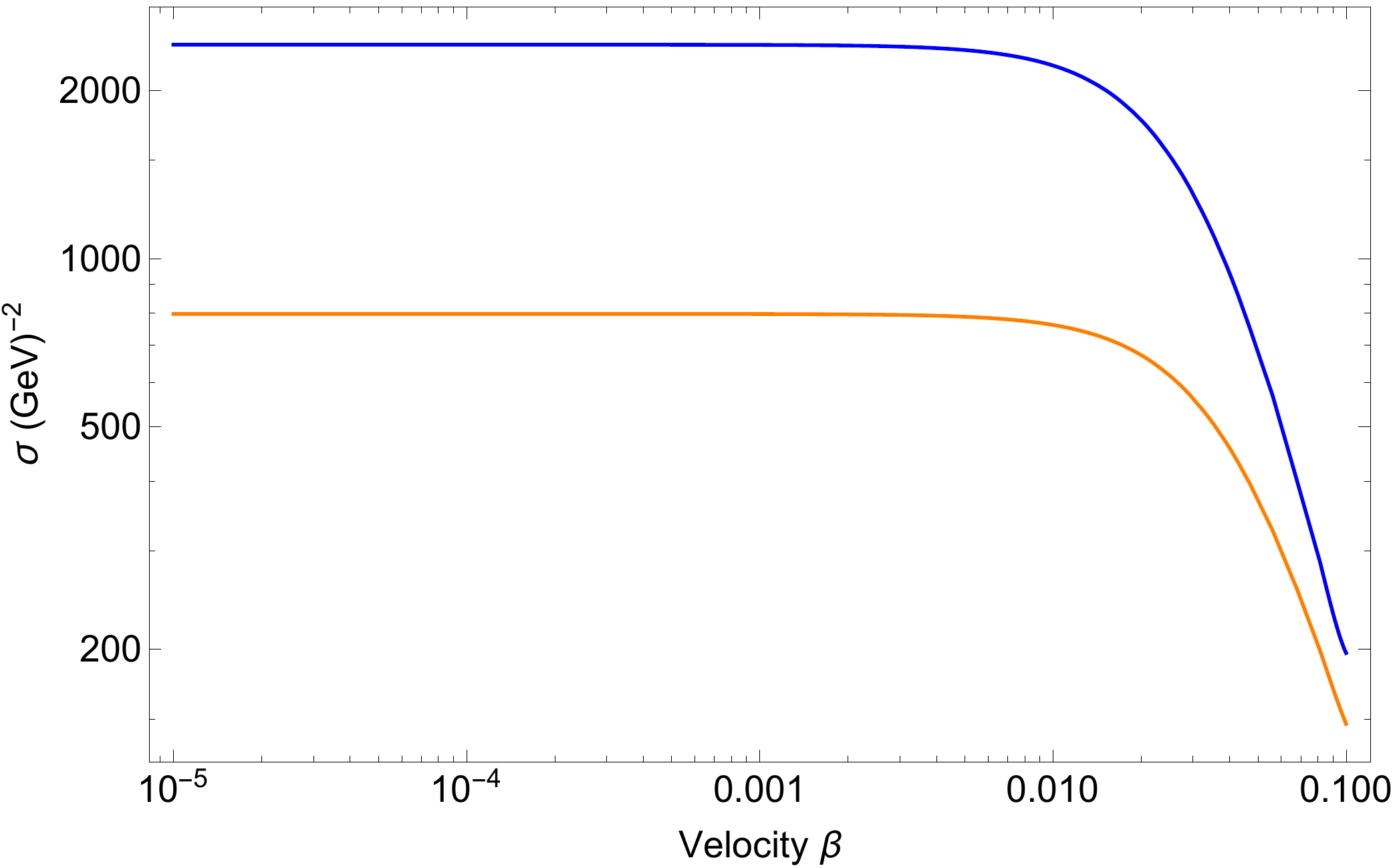}
    \caption{Cross section as a function of velocity for dark matter
    coupled via a scalar mediator. The numerical cross section (solid
  blue) is compared with the tree-level QFT cross section (solid
orange). The numerical cross sections shown here includes the
nonperturbative Sommerfeld enhancement and is summed over partial
waves (truncated at $\ell_{\text{max}}$ such that
$\sigma_{\ell_{\text{max}}} \leq 10^{-4}\sigma_{0}$). Parameters are
$\lambda = 1$, $m_{\chi} = 1$ GeV and $m_{\phi} = 10^{-1}$ GeV. At low
velocities, we observe a significant Sommerfeld enhancement but at
larger velocities, the numerical cross section becomes well
approximated by the Born cross section as expected.}
    \label{fig:Yukawa_Sommerfeld}
\end{figure}

\subsection{Yukawa Potential}
In Figure~\ref{fig:Yukawa_Sommerfeld}, we present our results for the
Yukawa potential generated via a scalar interaction. The solid orange
curve is the tree-level QFT cross section.  Using the procedure
outlined in Section~\ref{Boundary_Conditions}, we set the boundary
conditions and solve the Schr\"odinger equation. The numerical cross
section we obtain is plotted as the solid blue curve. We introduce a
dynamical cutoff to compute the numerical cross section. In principle,
the total cross section, $\sigma$, is given by summing the partial
wave cross sections, $\sigma_{\ell}$, for all the partial waves. In
practice, we truncate this sum at $\ell_{\text{max}}$ such that
$\sigma_{\ell_{\text{max}}} \leq 10^{-4}\sigma_{0}$. 
It is worth noting here that the Yukawa potential is well-behaved at
the origin and does not require the matching procedure. We checked
that the cross sections calculated with and without using our matching
procedure agree. For consistency with the rest of the results shown in
this paper, we report numerical cross sections for the Yukawa
potential computed using the matching procedure.
The results in Figure~\ref{fig:Yukawa_Sommerfeld} show that there is a
considerable enhancement in the cross section in the nonrelativistic
regime. At higher velocities, the QFT tree-level cross section becomes
a better approximation for the cross section. As has been discussed
previously in the literature, we expect Sommerfeld enhancement to be a
significant effect in the nonrelativistic limit for light mediators
and our results are in good agreement with this expectation. 

\subsection{Pseudoscalar Mediator}
\begin{figure}[tp]
    \centering
    \includegraphics[width=0.75\textwidth]{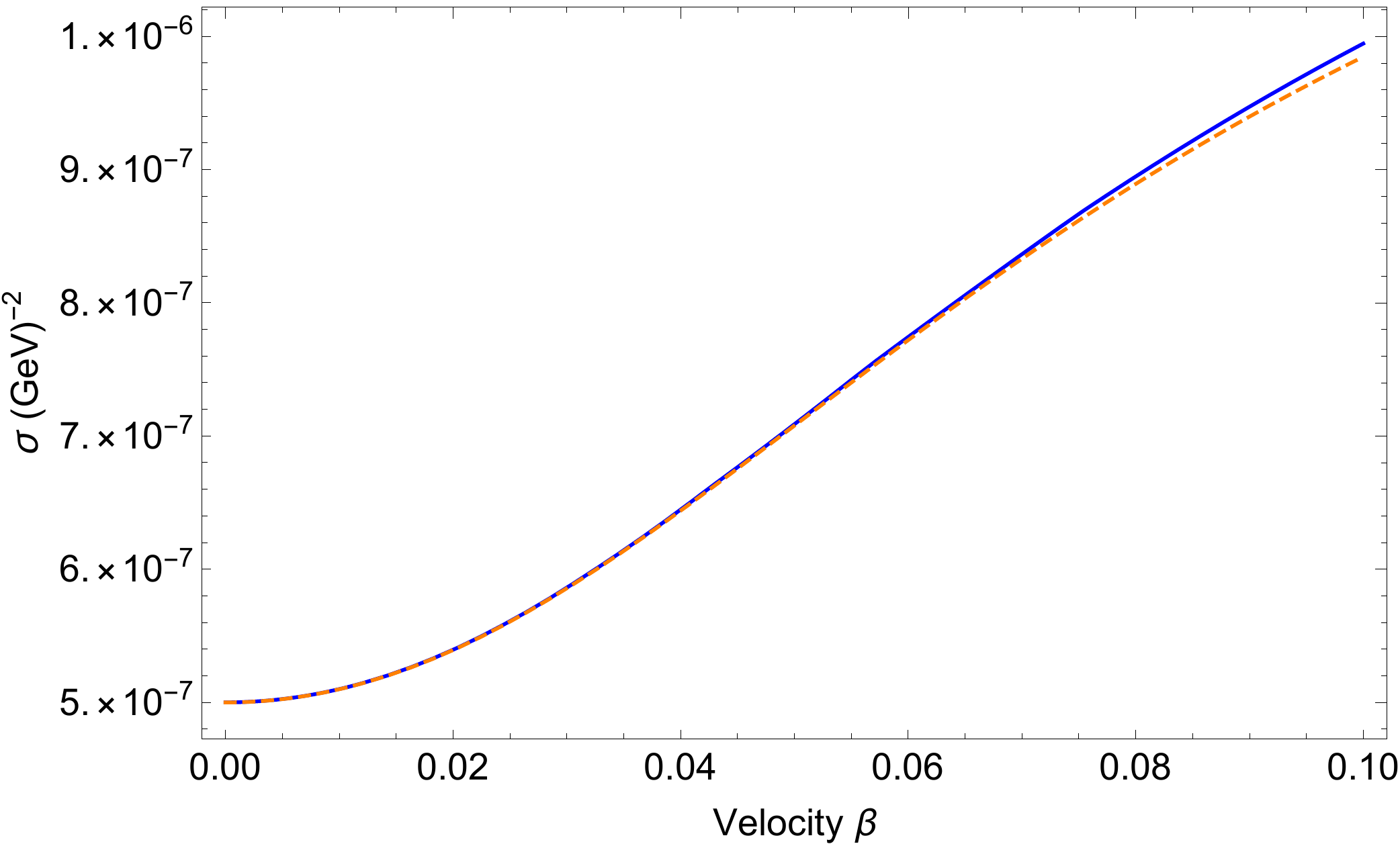}
    \caption{Cross section as a function of velocity for dark matter coupled via a pseudoscalar mediator. The numerical cross section (solid blue) computed with the procedure outlined in Section~\ref{gen_procedure} is compared with the tree-level QFT cross section (dashed orange). We set $\lambda = 10^{-1}$, $m_{\chi} = 1$ GeV and $m_{\phi} = 10^{-1}$ GeV. At low velocities we do not see any Sommerfeld enhancement. We begin to see deviations at larger velocities, as expected since the tree-level QFT answer is a fully relativistic calculation but the numerical cross section is determined from a nonrelativistic potential.}
    \label{fig:Pseudoscalar_Sommerfeld}
\end{figure}

Previously, in Figure $\ref{fig:QFT_Born_comparison}$ we showed the
agreement between the first Born approximation and the QFT expectation
for the cross section for a pseudoscalar mediator. 
Having set the boundary conditions, we compute the numerical cross
section, shown in blue, and compare it to the tree-level QFT cross
section in dashed orange in Figure
$\ref{fig:Pseudoscalar_Sommerfeld}$. We implement the same dynamical
cutoff on the sum over partial waves for the pseudoscalar potential as
we did for the Yukawa potential.
At large velocities, the two curves deviate by $\mathcal{O}(1\%)$.
This discrepancy is saturated by the nonrelativistic corrections which
scale as $\beta^{2}$ and we see that there is no additional
enhancement.  While we might have expected our intuition from the
Yukawa potential to apply here, we do not find any significant
Sommerfeld enhancement for any values of $m_{\phi}$.

The absence of enhancements in the small $m_{\phi}$ regime can be
understood intuitively. The pseudoscalar potential, in Equation
$\ref{pseudoscalar_potential_equation}$, has two Yukawa like terms
which in principle can generate an enhancement, but both of these
terms are suppressed by $\frac{m_{\phi}^{2}}{m_{\chi}^{2}}$.
Therefore, in the limit of small $m_{\phi}$, the suppression of these
terms shuts off possible Sommerfeld enhancement. A similar intuitive
argument is provided in \cite{Kahlhoefer:2017umn}.
Furthermore, the more divergent terms, such as the $r^{-3}$ term,
behave effectively like a short range potential and hence don't
generate enhancements. In the large $m_{\phi}$ limit, we don't expect
Sommerfeld enhancement in either the Yukawa or pseudoscalar case as
the exponential suppression takes over and we have short range
potentials.

\subsubsection{Diagrammatic Argument}
Feynman diagrammatic arguments lend further credence to this result.
We discuss the general idea and results of this argument for the
scalar and pseudoscalar case here and relegate the details to Appendix
\ref{Feynman_Diagrammatic_Argument}. Sommerfeld enhancement occurs in
a region of phase space where diagrams with two mediator exchanges are
comparable or parametrically larger than a tree level diagram with a
single mediator exchange. To understand the scaling behavior in this
regime, we can analytically compute the resulting box diagram for the
cases of scalar and pseudoscalar mediators.

We study the ratio of $\mathcal{M}_{\text{1-loop}}$ to
$\mathcal{M}_\text{tree}$. For the pseudoscalar case, the leading
behavior goes like $\frac{g^{2}}{16\cpi^2}\log\xi$, where we have
defined $\xi = m_\chi^2/m_\phi^2$.  On the other hand, for the scalar
case the leading behavior goes like
$\frac{g^{2}}{4\cpi }\sqrt{\xi}$. The dependence on the coupling $g$
is as expected. As  always for a  perturbative calculation, if we make
the coupling arbitrarily large, then loops will be important. What is
relevant for the existence of nonrelativistic enhancement is the
scaling with $\xi$, which is different for the scalar and pseudoscalar
cases. At large $\xi$, the ratio in the scalar case diverges, whereas
in the pseudoscalar case, it behaves like a log-enhanced loop factor.
This indicates that when we make our mediator light, the box diagram
becomes important and dominates over the tree level diagram in the
scalar, or more generally the Yukawa case. This is Sommerfeld
enhancement. It requires that we resum an infinite family of diagrams,
which is accomplished by solving the nonrelativistic Schr\"odinger
equation. On the other hand, in the pseudoscalar case, the box diagram
is not enhanced by nonrelativistic effects as the mediator becomes
light, which corroborates the numerical results and heuristic
arguments we gave above.

Our conclusions deviate from certain claims in the literature. Our
analysis of the EFT and its operators largely agrees with  that of
\cite{Bellazzini:2013foa}, though they included only $t$-channel
contributions from light mediators and neglected contact terms. Both
\cite{Bellazzini:2013foa} and \cite{Bedaque:2009ri} correctly explain
how pseudoscalar exchange can couple modes with $l = j \pm 1$.
However, \cite{Bellazzini:2013foa} then argued that scattering should
be dominated by a singular $1/r^3$ potential and proceeded to analyze
a single-channel equation with such a potential as an idealization of
the pseudoscalar exchange potential. We do not believe that this toy
single-channel problem has similar physics to pseudoscalar exchange.
In particular, our matching procedure works well for the pseudoscalar
exchange problem but is not even well-defined for the $l = 0$ mode in
a $1/r^3$ potential, because the integral computing the first Born
approximation diverges at small $r$. This reflects the fact that a
$1/r^3$ potential is singular whereas, as we have argued, a
$[3(S_{1}\cdot\hat{r})(S_{2}\cdot\hat{r}) - S_{1}\cdot S_{2}]/r^3$
potential is not (at least at leading order in the Born
approximation). The authors of \cite{Bedaque:2009ri} studied the
coupled-channel problem for $j = 1$ modes and concluded that there can
be significant Sommerfeld enhancement. Their approach is similar to
ours: they have introduced a short-distance square well regulator, and
then solve the full coupled-channel problem at longer distances.
However, they have not carried out a detailed matching to perturbative
QFT, and as a result they treat the depth of the square well potential
and the coupling strength as free parameters to vary. The Sommerfeld
enhancement that they observe arises in regions of nonperturbatively
large short-distance scattering. We believe that  there is no conflict
between their numerical  results and our claim that Sommerfeld
enhancement does not arise within the parameter space one can obtain
through perturbative matching to a weakly-coupled quantum field
theory.

\subsection{Axial Vector Mediator}
As discussed in Section~\ref{Explicit_Examples}, the axial vector
interaction is non-trivial. It comes along with a radial Higgs mode
that also couples to the fermions. This is a scalar particle coupled
to our fermionic dark matter, which in turn will induce a Yukawa
potential between them. As we saw already, Yukawa potentials generate
a Sommerfeld enhancement. On the other hand, if we want to study just
the axial vector type interaction, we need to decouple the Higgs mode
by making it heavier. One consistent way of achieving this is by
making the square root of the Higgs quartic larger than the Higgs
coupling to the dark matter ($\sqrt{2\lambda_{q}} > y$). Furthermore,
to make the axial vector mediator lighter than the dark matter, we
need the gauge coupling to be smaller than the Higgs-dark matter
coupling ($y > \sqrt{8}\lambda$). By doing so, we can decouple the
Higgs and study a pure axial vector theory. Formally, by integrating
out the Higgs, we generate a four-Fermi interaction which manifests as
a delta function in the potential since it was generated by a contact
interaction without any ${\vec q}$ dependence. 

Having done so, we can now isolate the potential generated by a purely
axial vector type interaction. By inspecting Equation
$\ref{axial_vector_potential_equation}$, we see a term proportional to
$V_{\text{pseudoscalar}}$. Again, this term does not generate a
Sommerfeld enhancement. In addition, there is a Yukawa term in the
potential. This term does not generate a Sommerfeld enhancement
because in the light mediator regime, the coupling, which is
proportional to the mediator mass, is also small. On the other hand,
if we want a large coupling, then the mediator mass increases
proportionally and Sommerfeld enhancement turns off. The structure of
the underlying UV completion restricts the parameter choices we can
make. We checked that the results obtained from the QFT match the
results obtained from the full quantum mechanical calculation for the
axial vector potential when working in the decoupling limit, which
supports the heuristic argument we make above. Therefore, when working
in this limit, we can also compute $\langle\sigma_Vv\rangle$ using the
relativistic, perturbative QFT. 

\subsection{Results}
\label{sec:results}

\begin{figure}[tp]
    \centering
    \includegraphics[width=0.75\textwidth]{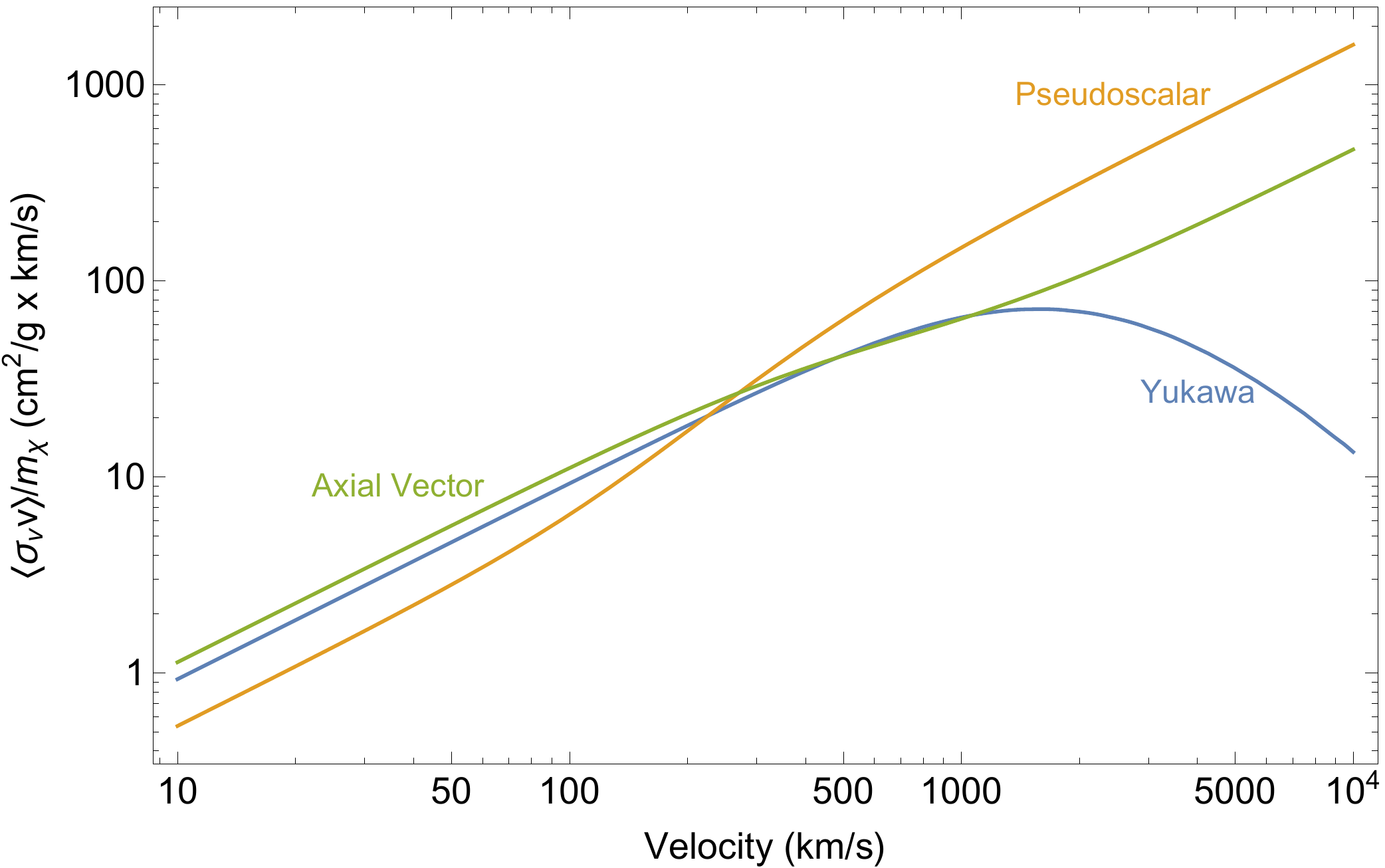}
    \caption{Velocity weighted viscosity cross section as a function of average velocity for dark matter coupled via a scalar, pseudoscalar or an axial vector mediator. For the scalar case, we choose $\lambda = 10^{-1}$, $m_{\phi} = 10^{-2}$ GeV and $m_{\chi} = 1$ GeV. For the pseudoscalar case, we choose $\lambda = 10^{-1}$, $m_{\phi} = 10^{-3}$ MeV and $m_{\chi} = 1$ MeV. For the axial vector case, we work in the decoupling limit. We choose the vev $\langle v \rangle = 20$ MeV, $\lambda_{\text{q}} = 4\cpi$, $\lambda = 10^{-3}$ and $y = 1$. We assume the dark matter follows a Maxwell-Boltzmann distribution and use a hard cutoff at the escape velocity. The full numerical cross section for the Yukawa potential includes the nonperturbative Sommerfeld enhancement. On the other hand, as we discuss in the text, the axial vector interaction in the decoupling limit and the pseudoscalar interaction don't induce Sommerfeld enhancement and are therefore computed using tree-level relativistic QFT.}
    \label{fig:sigmav}
\end{figure}

In this section, we show the results from the procedure laid out in Section~\ref{Boundary_Conditions}. 
The quantity of merit is the viscosity cross section $\sigma_V$. The angular weighting regulates both forward and backward scattering, which is important since singularities in the forward and backward scattering limit, although physical, do not change the dark matter velocity distribution and hence have no observable effect. The physical quantity we can extract from measurements is $\langle\sigma_Vv\rangle$. The velocity averaging assumes a Maxwell-Boltzmann distribution truncated at an escape velocity $v_\text{esc} = \sqrt{2} v_\text{rms}$, as for a virialized halo. The results are shown in Figure~\ref{fig:sigmav} for the various interactions we considered. 
The parameters are chosen such that the cross sections are at approximately the correct order of magnitude over the velocity range of interest. A more dedicated exploration of the viable parameter space fitting the self-interaction cross section measurements from astrophysical data in \cite{Kaplinghat:2015aga} should be performed, but is beyond the scope of this work.

As we discussed in detail above, our numerical results for the pseudoscalar mediator and the axial vector mediator, with a decoupled Higgs, show no Sommerfeld enhancement. So, we can compute $\langle\sigma_Vv\rangle$ directly from perturbative, relativistic QFT in both of these cases. 
For the pseudoscalar mediator, we notice a kink occurring at $v \sim \mathcal{O}(m_{\phi}/m_{\chi})$. The kink exhibits a characteristic factor of 3 increase in the cross section. The cross section plateaus before and after the kink. These are robust predictions for the behavior of the pseudoscalar interaction. The features we highlighted above can be seen more clearly in Figure~\ref{fig:QFT_Born_comparison}, even though it contains only the tree-level QFT results, since these are a good approximation to the answer in the case of pseudoscalar and axial vector interactions.

\section{Conclusions}
\label{sec:conclusions}

In this work, we have studied the velocity dependence of interactions between spin-$1/2$ dark matter particles mediated by a light boson. In particular, we studied the scenario where the boson is a scalar, vector, pseudoscalar or axial vector. We derived the associated potentials including both $s$- and $t$-channel contributions to the scattering process.  
We outlined a new procedure for setting the boundary conditions where we match a tree-level perturbative QFT estimate of short-distance scattering to the Born approximation to short-distance scattering in the effective nonrelativistic quantum mechanical theory. Numerically solving the Schr\"odinger equation then allows us to capture the effect of Sommerfeld enhancement. While we have only considered simplified models in this work, our procedure generalizes straightforwardly to more complicated models as well. 

We presented numerical results for the scalar and pseudoscalar case. The scalar mediator generates significant enhancement for low velocities and light mediators, an effect that has been studied extensively in the literature previously. Our numerical results for the pseudoscalar mediator show an excellent match to the tree-level perturbative QFT approximation. This lack of Sommerfeld enhancement is further supported by a Feynman diagrammatic argument. 
We also argued that Sommerfeld enhancement is absent for an axial vector mediator. In this scenario, the mediator mass and the gauge coupling are tied together such that as the mass is dialed down, the coupling gets correspondingly weaker. 
Our results suggest that, if the shape of the cross section discussed  in  \cite{Kaplinghat:2015aga} persists with more data and a  better understanding of baryonic effects, then pseudoscalar and axial mediators will not fit the data as well as scalar and vector mediators. (These light-mediator models are not the only options, however; see, e.g., \cite{Chu:2018faw,Chu:2019awd,Chu:2018fzy}).

The matching procedure that we have described can be applied beyond the four simple models we have studied. For example, the axial vector model also in general has scattering mediated by the Higgs boson that provides a mass to the axial vector field. One could match  to a theory that includes both the Higgs and axial-vector contributions. In general, our matching procedure will be useful in cases with both long-range Sommerfeld enhanced scattering and short-distance contributions to the amplitude. Once the velocity-dependence of the cross section in a given model is known, it can be fed into simulations or other studies of structure formation, for instance in the ETHOS framework \cite{Cyr-Racine:2015ihg,Vogelsberger:2015gpr,Rivero:2018bcd}.

Our results also raise a more abstract question: to what extent are singular potentials in quantum mechanics relevant when matching  to an underlying perturbative QFT? There is a large literature on singular quantum-mechanical potentials like $1/r^3$. We have observed that when matching to tree-level QFT, such  terms  are accompanied by spin-dependent factors that eliminate the dangerous terms in the leading-order Born approximation. In the future, it would be interesting to better understand the general  properties of quantum mechanical models arising from weakly coupled QFTs.

\acknowledgments{We thank Pouya Asadi and Sruthi A. Narayanan for interesting discussions. 
 PA is supported by NSF grants PHY-1620806 and PHY-1915071, the Chau Foundation
HS Chau postdoc support award, the Kavli Foundation grant Kavli Dream Team, and the
Moore Foundation Award 8342.
AP is supported in part by an NSF Graduate Research Fellowship grant DGE1745303 and by the DOE Grant DE-SC0013607. MR is supported in part by the NASA ATP Grant NNX16AI12G and by the DOE Grant DE-SC0013607. A portion of this work was completed at the KITP, supported in part by the National Science Foundation grant PHY05-51164,
at the Munich Institute for Astro- and Particle Physics (MIAPP) which is funded by the Deutsche Forschungsgemeinschaft (DFG, German Research Foundation) under Germany's Excellence Strategy -- EXC-2094 -- 390783311,
and at the Aspen Center for Physics, supported by National Science Foundation grant PHY-1607611. For checking some of our analytic results regarding matrix elements in different angular momentum bases we have found the \href{http://homepage.cem.itesm.mx/lgomez/quantum/}{Quantum Mathematica package} by Jos{\'e} Luis G{\'o}mez-Mu{\~n}oz and Francisco Delgado useful.
}
\appendix
\section{Angular Momentum Basis Conversions}
\label{Pseudoscalar_Clebsch_Gordan}

The potentials we are looking at are spherically symmetric potentials. For the case of scalar and vector interactions, the following decomposition is unnecessary, but for the potentials generated by pseudoscalar and axial vector interactions, we have to evaluate the matrix elements of operators such as $S_{1}\cdot S_{2}$ and $3(S_{1}\cdot\hat{r})(S_{2}\cdot\hat{r}) - S_{1}\cdot S_{2}$. These operators can change spin and orbital angular momentum but, due to the overall spherical symmetry, won't change total angular momentum $J^{2}$ or $J_{z}$. So, we can consider our states to be labeled by the eigenvalues of  $J^{2}, L^{2}, S^{2}$, and $J_{z}$ and look at the decomposition of one of these states into the basis of states labelled by $L^{2}, L_{z}, S^{2}, S_{z}$. 

Consider a state with eigenvalues $j(j+1)$ and $\sigma$ for $J^{2}$ and $J_{z}$ respectively. For spin-1/2 fermions the total spin can either be 0 or 1, as we saw above. The state with the spin singlet configuration looks like
\begin{equation}
|j,\sigma,j,0 \rangle_{j,m_{j},l,s} = |j,\sigma \rangle_{l,m_{l}} \otimes |0,0 \rangle_{s,m_{s}}
\end{equation}
The states with the spin triplet configurations look like
\begin{equation}
\begin{split}
|j,\sigma,j+1,1 \rangle_{j,m_{j},l,s} = &\sqrt{\frac{(j-\sigma+1)(j-\sigma+2)}{2(j+1)(2j+3)}}|j+1,\sigma - 1 \rangle_{l,m_{l}} \otimes |1,1 \rangle_{s,m_{s}} \\ - &\sqrt{\frac{(j-\sigma+1)(j+\sigma+1)}{(j+1)(2j+3)}}|j+1,\sigma \rangle_{l,m_{l}} \otimes |1,0 \rangle_{s,m_{s}} \\ + &\sqrt{\frac{(j+\sigma+1)(j+\sigma+2)}{2(j+1)(2j+3)}}|j+1,\sigma + 1 \rangle_{l,m_{l}} \otimes |1,-1 \rangle_{s,m_{s}}
\end{split}
\end{equation}

\begin{equation}
\begin{split}
|j,\sigma,j,1 \rangle_{j,m_{j},l,s} = &-\sqrt{\frac{(j-\sigma+1)(j+\sigma)}{2j(j+1)}}|j,\sigma - 1 \rangle_{l,m_{l}} \otimes |1,1 \rangle_{s,m_{s}} \\ + &\sqrt{\frac{\sigma^{2}}{j(j+1)}}|j,\sigma \rangle_{l,m_{l}} \otimes |1,0 \rangle_{s,m_{s}} \\ + &\sqrt{\frac{(j-\sigma)(j+\sigma+1)}{2j(j+1)}}|j,\sigma + 1 \rangle_{l,m_{l}} \otimes |1,-1 \rangle_{s,m_{s}}
\end{split}
\end{equation}

\begin{equation}
\begin{split}
|j,\sigma,j-1,1 \rangle_{j,m_{j},l,s} = &\sqrt{\frac{(j+\sigma-1)(j+\sigma)}{2j(2j-1)}}|j-1,\sigma - 1 \rangle_{l,m_{l}} \otimes |1,1 \rangle_{s,m_{s}} \\ + &\sqrt{\frac{(j-\sigma)(j+\sigma)}{j(2j-1)}}|j-1,\sigma \rangle_{l,m_{l}} \otimes |1,0 \rangle_{s,m_{s}} \\ + &\sqrt{\frac{(j-\sigma-1)(j-\sigma)}{2j(2j-1)}}|j-1,\sigma + 1 \rangle_{l,m_{l}} \otimes |1,-1 \rangle_{s,m_{s}}
\end{split}
\end{equation}
\subsection{$\vec{S}\cdot\hat{r}$ Operator}
A useful decomposition of $\vec{S}\cdot\hat{r}$ is given by
\begin{equation}
\vec{S}\cdot\hat{r} = S_{x} \sin\theta \cos\phi + S_{y}\sin\theta \sin\phi + S_{z}\cos\theta = \frac{1}{2}[S_{+}\text{e}^{-\text{i}\phi}\sin\theta + S_{-}\text{e}^{\text{i}\phi}\sin\theta] + S_{z}\cos\theta
\end{equation}
The action of the angular operators is given by
\begin{equation}
(\cos \theta) \cdot Y^m_\ell(\theta, \phi) = \sqrt\frac{(\ell+1+m)(\ell+1-m)}{(2\ell+1)(2\ell+3)} Y^m_{\ell+1}(\theta,\phi) + \sqrt\frac{(\ell+m)(\ell-m)}{(2\ell+1)(2\ell-1)} Y^m_{\ell-1}(\theta,\phi)
\end{equation}

\begin{equation}
({\rm e}^{{\rm i}\phi} \sin \theta) \cdot Y^m_\ell(\theta, \phi) = -\sqrt\frac{(\ell+m+2)(\ell+m+1)}{(2\ell+1)(2\ell+3)} Y^{m+1}_{\ell+1}(\theta,\phi) + \sqrt\frac{(\ell-m)(\ell-m-1)}{(2\ell+1)(2\ell-1)} Y^{m+1}_{\ell-1}(\theta,\phi)
\end{equation}

\begin{equation}
({\rm e}^{-{\rm i}\phi} \sin \theta) \cdot Y^m_\ell(\theta, \phi) = \sqrt\frac{(\ell-m+2)(\ell-m+1)}{(2\ell+1)(2\ell+3)} Y^{m-1}_{\ell+1}(\theta,\phi) - \sqrt\frac{(\ell+m)(\ell+m-1)}{(2\ell+1)(2\ell-1)} Y^{m-1}_{\ell-1}(\theta,\phi)
\end{equation}

The action of $\mathcal{O}_{T} \equiv 3(\vec{S_{1}}\cdot\hat{r})(\vec{S_{2}}\cdot\hat{r}) - \vec{S}_{1}\cdot\vec{S}_{2}$ on our states is

\begin{equation}
\mathcal{O}_{T}|j,\sigma,j,0\rangle_{j,m_{j},l,s} = 0
\end{equation}

\begin{equation}
\mathcal{O}_{T}|j,\sigma,j,1\rangle_{j,m_{j},l,s} = \frac{1}{2}|j,\sigma,j,1\rangle_{j,m_{j},l,s}
\end{equation}

\begin{equation}
\mathcal{O}_{T}|j,\sigma,j+1,1\rangle_{j,m_{j},l,s} = \frac{-(j+2)}{2(2j+1)}|j,\sigma,j+1,1\rangle_{j,m_{j},l,s} + \frac{3\sqrt{j(j+1)}}{2(2j+1)}|j,\sigma,j-1,1\rangle_{j,m_{j},l,s}
\end{equation}

\begin{equation}
\mathcal{O}_{T}|j,\sigma,j-1,1\rangle_{j,m_{j},l,s} = \frac{-(j-1)}{2(2j+1)}|j,\sigma,j-1,1\rangle_{j,m_{j},l,s} + \frac{3\sqrt{j(j+1)}}{2(2j+1)}|j,\sigma,j+1,1\rangle_{j,m_{j},l,s}
\end{equation}
So we see that the states with angular momentum $j+1$ and $j-1$ mix with each under the action of the operator $\mathcal{O}_{T}$.

\begin{equation}
\mathcal{O}_{T}\begin{pmatrix}
|j,\sigma,j+1,1\rangle_{j,m_{j},l,s} \\[6 pt]
|j,\sigma,j-1,1\rangle_{j,m_{j},l,s}
\end{pmatrix} = 
\begin{pmatrix}
\frac{-(j+2)}{2(2j+1)} & \frac{3\sqrt{j(j+1)}}{2(2j+1)} \\[6 pt]
\frac{3\sqrt{j(j+1)}}{2(2j+1)} & \frac{-(j-1)}{2(2j+1)} \\
\end{pmatrix}
\begin{pmatrix}
|j,\sigma,j+1,1\rangle_{j,m_{j},l,s} \\[6 pt]
|j,\sigma,j-1,1\rangle_{j,m_{j},l,s}
\end{pmatrix}
\end{equation}

\section{Fermion/Antifermion Spin Matrices and Minus Signs}
\label{Spinor_Minus_Signs}
Following the conventions of \cite{Peskin:1995ev}, we have the following definitions:
\begin{equation}
J_{z} a_{0}^{s\dagger} | 0 \rangle = \pm \frac{1}{2} a_{0}^{s\dagger} | 0 \rangle \quad \quad J_{z} b_{0}^{s\dagger} | 0 \rangle = \mp \frac{1}{2} b_{0}^{s\dagger} | 0 \rangle
\end{equation}
\begin{equation}
\text{Upper Sign} \quad  \xi^{s} = \begin{pmatrix} 1 \\ 0 \end{pmatrix} \quad \quad \text{Lower Sign} \quad \xi^{s} = \begin{pmatrix} 0 \\ 1 \end{pmatrix}
\end{equation}
This means that for a particle, we have
\begin{equation}
\xi^{s\dagger}\sigma_{z}\xi^{s'} = \pm \xi^{s\dagger}\xi^{s'} = \pm \delta^{ss'}
\end{equation}
and for an antiparticle, we have
\begin{equation}
\xi^{s\dagger}\sigma_{z}\xi^{s'} = \mp \xi^{s\dagger}\xi^{s'} = \mp \delta^{ss'}
\end{equation}
More generically, let's identify $\eta = \epsilon\xi_{\bar{\chi}}^{*}$. $\epsilon$ is the antisymmetric tensor and $\sigma$ satisfies the relation: $\sigma^{i}\epsilon = -\epsilon(\sigma^{i})^{*}$.
\begin{equation}
\eta^{\dagger}\vec{\sigma}\eta = (\epsilon\xi_{\bar{\chi}}^{*})^{\dagger}\vec{\sigma}(\epsilon\xi_{\bar{\chi}}^{*}) = \xi_{\bar{\chi}}^{T}\epsilon\vec{\sigma}\epsilon\xi_{\bar{\chi}}^{*} = \xi_{\bar{\chi}}^{T}(-\vec{\sigma}^{*})\xi_{\bar{\chi}}^{*}
\end{equation}
Now this is a scalar quantity so we are free to transpose it and this gives us $-\xi_{\bar{\chi}}^{\dagger}\vec{\sigma}\xi_{\bar{\chi}}$, where this $\vec{\sigma}$ gets identified with the spin matrix $\vec{S}$. From this, we see that the minus sign naturally arises when looking at $\eta^{\dagger}\vec{\sigma}\eta$ for any generic state, and not just the $z$ eigenstates.

\section{Feynman Diagrammatic Argument for Sommerfeld Enhancement}
\label{Feynman_Diagrammatic_Argument}
\begin{figure}[h!]
\begin{center}
    \includegraphics[width=0.5\textwidth]{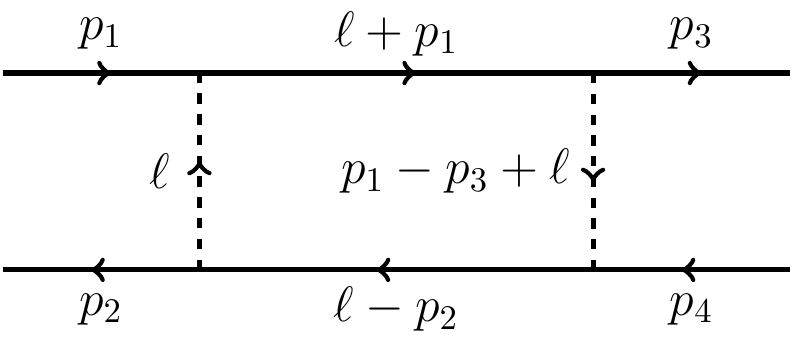}
\end{center}
\caption{The box diagram corresponds to the first diagram in the infinite set of ladder diagrams being resummed by our procedure. Sommerfeld enhancement arises when this diagram gives a contribution that is comparable to or larger than the tree level contribution to the scattering process.}
\label{fig:box_diagram}
\end{figure}

There are a few lines of evidence supporting our claim that Sommerfeld enhancement is absent in the pseudoscalar case. We can show this analytically by computing the box diagram in Figure $\ref{fig:box_diagram}$. The general amplitude for this diagram is given by
\begin{align}
\mathcal{M}_\text{1-loop}
&\sim 
g^4\int \frac{{\rm d}^4 l}{(2\cpi)^4} 
\frac{\mathcal{N}}
{
[(p_{1}+l)^{2}-m_{\chi}^{2}]
[(l-p_{2})^{2}-m_{\chi}^{2}]
[l^{2}-m_{\phi}^{2}]
[(p_{1}-p_{3}+l)^{2}-m_{\phi}^{2}]}
\end{align}
where the numerator $\mathcal{N}$ is
\begin{align}
    \mathcal{N}
    &=
    \overline{u}(p_{3})
    \left(
    \Gamma
    (\slashed{p}_1+\slashed{l}+m_{\chi})
    \Gamma
\right) 
u(p_{1})
\times
\overline{v}(p_{2})
\left(
\Gamma
(\slashed{l}-\slashed{p}_2+m_{\chi})
\Gamma\right) v(p_{4})
\end{align}
The $\Gamma$ matrices represent the matrix structure arising from the vertices. We will focus on two cases: the Yukawa interaction, where $\Gamma$ is the identity matrix, and the pseudoscalar case, where $\Gamma = \gamma^5$.

On the equations of motion, for the pseudoscalar case, the numerator simplifies to
\begin{equation}
\overline{u}(p_{3})\slashed{l}u(p_{1})\overline{v}(p_{2})\slashed{l}v(p_{4}).
\end{equation}
We introduce Feynman parameters and perform the integral over the loop momentum. Since we are interested in the nonrelativistic regime, we take the $v\rightarrow 0$ limit of the amplitude. This allows us to perform the integration over two of the Feynman parameters and we are left with the following expression. 
\begin{equation}
    \begin{split}
\mathcal{M}_\text{1-loop} \sim &-\iu  g^{4}\overline{u}(p_{3})\gamma_{\mu}u(p_{1})\overline{v}(p_{2})\gamma^{\mu}v(p_{4})\int_{0}^{1}{\rm d}w\int_{0}^{1-w}{\rm d}x
\frac{(1-w-x)}{32\cpi^{2}(m_{\chi}^{2}(w-x)^{2} + m_{\phi}^{2}(1-w-x))} \\ 
&+\iu g^{4}\overline{u}(p_{3})u(p_{1})\overline{v}(p_{2})v(p_{4})\int_{0}^{1}{\rm d}w\int_{0}^{1-w}{\rm d}x \frac{m_{\chi}^{2}(w-x)^{2}(1-w-x)}{16\cpi^{2}(m_{\chi}^{2}(w-x)^{2} + m_{\phi}^{2}(1-w-x))^{2}}.
\end{split}
\end{equation}
In the nonrelativistic limit, the leading term from the $\gamma$ matrices comes from the $\gamma^{0}$. We also define $\xi = m_{\chi}^{2}/m_{\phi}^{2}$. This allows us to combine the two terms into 
\begin{equation}
\mathcal{M}_\text{1-loop} \sim \frac{\iu g^{4}\overline{u}(p_{3})u(p_{1})\overline{v}(p_{2})v(p_{4})}{32\cpi^{2}m_{\phi}^{2}}\int_{0}^{1}{\rm d}w\int_{0}^{1-w}{\rm d}x \frac{\xi(w-x)^{2}(1-w-x)-(1-w-x)^{2}}{(\xi(w-x)^{2}+(1-w-x))^{2}}.
\label{eq:wx_integral}
\end{equation}
This integral can be computed analytically and we obtain the following result
\begin{equation}
\begin{split}
\frac{1}{2\xi^{2}\sqrt{-1+4\xi}}\Bigg((2-6\xi)\arctan\Big[\frac{1}{\sqrt{-1+4\xi}}\Big] + (-2+6\xi)\arctan\Big[\frac{1-2\xi}{\sqrt{-1+4\xi}}\Big] + \\
\sqrt{-1+4\xi}(2\xi + \log\xi - \xi\log\xi)\Bigg).
\end{split}
\end{equation}
A series expansion around large $\xi$ yields
\begin{equation}
\frac{1}{2}\Bigg(\frac{2-\log\xi}{\xi} - \frac{3}{2}\cpi\Big(\frac{1}{\xi}\Big)^{3/2} + \frac{3 + 2\log\xi}{2\xi^{2}} + \mathcal{O}\Big(\frac{1}{\xi}\Big)^{5/2}\Bigg).
\end{equation}
The box diagram then gives a contribution to the matrix element scaling as
\begin{equation}
{\cal M}_\text{1-loop} \sim \frac{g^4}{32\cpi^2} \log\frac{m_\chi^2}{m_\phi^2}.
\end{equation}
This should be compared with the tree-level amplitude, which scales as ${\cal M}_\text{tree} \sim g^2$. Note that although the $t$-channel contribution to the tree-level amplitude is momentum suppressed in the nonrelativistic limit, the $s$-channel contribution is not. Hence ${\cal M}_\text{1-loop}/{\cal M}_\text{tree}$ is on the order of a naive, log-enhanced loop factor, and nonrelativistic effects do not enhance the cross section predicted by perturbative QFT.

By contrast, for the scalar case, we expect to find a Sommerfeld enhancement. The numerator simplifies to\footnote{In the vector case, a priori the $\Gamma$ should be $\gamma_{\mu}$, but in the nonrelativistic limit, the dominant contribution comes from $\gamma_{0}$. Therefore, even the vector case maps back down to the scalar case and the argument goes through in the same manner which is why we get Sommerfeld enhancement for an attractive Yukawa potential which can be generated by scalar and vector mediators.}
\begin{align}
\mathcal{N}
&=
\overline{u}(p_{3})(\slashed{l}+2m_{\chi})u(p_{1})\overline{v}(p_{2})(\slashed{l}+2m_{\chi})v(p_{4}).
\end{align}
Computing the integral over Feynman parameters analytically for the scalar case yields
\begin{equation}
\begin{split}
-\frac{1}{2\xi^{2}\sqrt{-1+4\xi}}\Bigg((-2+6\xi-32\xi^{3})\arctan\Big[\frac{1}{\sqrt{-1+4\xi}}\Big] + (2-6\xi+32\xi^{3})\arctan\Big[\frac{1-2\xi}{\sqrt{-1+4\xi}}\Big] + \\
\sqrt{-1+4\xi}(-2\xi - \log\xi + \xi\log\xi)\Bigg).
\end{split}
\end{equation}
For large $\xi$, this behaves like $4\cpi\sqrt{\xi}$. This means that the Sommerfeld enhancement can  be important for low mediator masses, when
\begin{equation}
\frac{\mathcal{M}_\text{1-loop}}{\mathcal{M}_\text{tree}} \gtrsim 1 \quad \Rightarrow \quad m_\phi \lesssim \frac{g^2 m_\chi}{4\cpi}.
\end{equation}
This is consistent with the standard claim about the regime of nonrelativistic enhancement for a Yukawa potential (e.g., \cite{Tulin:2013teo}).

One can understand the origin of the enhancement as follows: the numerator of the integral scales as $\xi$ and the denominator as $\xi^2$, so in most of the integration region one expects a suppressed contribution at large $\xi$. However, when $(w-x)^2 \sim \xi^{-1}$, the denominator takes order-one values and the integrand is of order $\xi$. This occurs only in a region of size $\xi^{-1/2}$ within the overall integration region, and accounts for an integral of size $\sqrt{\xi}$. This did not occur in the pseudoscalar  case, where one can check that, precisely when the denominator becomes of order one, the numerator is suppressed as well.

\bibliographystyle{JHEP.bst}
\bibliography{ref.bib}
\end{document}